\documentclass[a4paper,11pt]{article}
\usepackage{pos}

\newcommand{\TTRR}{\mathbb{T}^2\times\mathbb{R}^2}
\newcommand{\SRRR}{\mathbb{S}^1\times\mathbb{R}^3}

\title{Novel first-order phase transition and critical points on $SU(3)$ Yang-Mills theory in $\mathbb{T}^2\times\mathbb{R}^2$}
\ShortTitle{$SU(3)$ Yang-Mills theory on $\mathbb{T}^2\times\mathbb{R}^2$}

\author*[a,b]{Masakiyo Kitazawa}
\author[c,d]{Daisuke Fujii}
\author[d]{Akihiro Iwanaka}
\author[e]{Daiki Suenaga}

\affiliation[a]{Yukawa Institute for Theoretical Physics,
Kyoto University, \\
Kyoto 606-8502, Japan}

\affiliation[b]{J-PARC Branch, KEK Theory Center, 
  Institute of Particle and Nuclear Studies, KEK, \\ Tokai, Ibaraki 319-1106, Japan}

\affiliation[c]{Advanced Science Research Center, Japan Atomic Energy Agency (JAEA), \\
Tokai, 319-1195, Japan}

\affiliation[d]{Research Center for Nuclear Physics, Osaka University, \\
Ibaraki 567-0048, Japan}

\affiliation[e]{Kobayashi-Maskawa Institute for the Origin of Particles and the Universe,
Nagoya University, \\
Nagoya, 464-8602, Japan}

\emailAdd{kitazawa@yukawa.kyoto-u.ac.jp}

\abstract{We investigate the thermodynamics and phase structure of $SU(3)$ Yang-Mills theory on $\TTRR$ with anisotropic spatial volumes in Euclidean spacetime in lattice numerical simulations and an effective model. In lattice simulations, the energy-momentum tensor defined through the gradient flow is used for the analysis of the stress tensor on the lattice. It is found that a clear pressure anisotropy is observed only at a significantly shorter spatial extent compared with the free scalar theory. We then study the thermodynamics obtained on the lattice in an effective model that incorporates two Polyakov loops along two compactified directions as dynamical variables. The model is constructed to reproduce thermodynamics measured on the lattice. The model analysis indicates the existence of a novel first-order phase transition and critical points as its endpoints. We argue that the interplay of the Polyakov loops induces the first-order transition.}

\FullConference{The 41st International Symposium on Lattice Field Theory (LATTICE2024)\\
 28 July - 3 August 2024\\
Liverpool, UK\\}

\begin{document}
\begin{flushright}
YITP-25-19, J-PARC-TH-0312
\end{flushright}
\maketitle

\section{Introduction}

Boundary conditions (BCs) in quantum field theory provoke various interesting phenomena, such as the Casimir effect in quantum electrodynamics~\cite{Casimir:1948dh}. 
Because the temperature $T$ is introduced as the BC along the imaginary-time direction in the Euclidean spacetime in the Matsubara formalism for thermal field theory, phenomena in thermal systems can also be regarded as those arising from BCs.

In this proceeding, we investigate $SU(N)$ Yang-Mills (YM) theory at nonzero temperature with the periodic boundary condition (PBC) along one spatial direction~\cite{Kitazawa:2019otp,Suenaga:2022rjk,Fujii:2024llh}. To be specific, we choose $x$ axis for the PBC and denote the length to this direction as $L_x$. Since thermal systems for bosons have a PBC along the imaginary-time direction of length $L_\tau=1/T$, this system is defined on a manifold $\TTRR$ in Euclidean spacetime with two PBCs of length $L_\tau$ and $L_x$.

YM theory on $\TTRR$ can have non-trivial phenomena compared to that in the thermodynamic limit on $\SRRR$. First, the length $L_x$ plays the role of a new external thermal parameter controlling the system. Second, since the rotational symmetry is broken by the PBC in the $x$ direction, the pressure becomes anisotropic. This introduces a new thermodynamics variable into the system. Finally, the YM theory on $\TTRR$ has two $Z_N$ symmetries corresponding to the twist at each BC, both of which can be spontaneously broken with the variation of external parameters.

In this proceeding, we first study this system for $N=3$ in lattice Monte Carlo simulations~\cite{Kitazawa:2019otp}. To investigate the anisotropic pressure, we measure the expectation values of energy-momentum tensor defined through the technique developed in Refs.~\cite{Suzuki:2013gza,Kitazawa:2016dsl,Iritani:2018idk} based on the gradient flow~\cite{Luscher:2010iy}. We show that the thermodynamics of $SU(3)$ YM theory near but above the critical temperature of deconfinement phase transition ($T_c$) is remarkably insensitive to the PBC compared with the massless free theory.
We then explore these results in an effective model to gain physical insights into the lattice results~\cite{Suenaga:2022rjk,Fujii:2024llh}. Aiming to describe spontaneous symmetry breakings of two $Z_N$ symmetries on $\TTRR$, we propose an effective model containing two Polyakov loops along two compactified directions as dynamical variables. We show that the model analysis indicates a possible existence of a novel first-order phase transition on $\TTRR$ in the deconfined phase, which terminates at critical points that should belong to the two-dimensional $Z_2$ universality class.

\section{Lattice Study}
\label{sec:lattice}

We have performed numerical simulations of $SU(3)$ YM
theory on four-dimensional Euclidean lattices with the standard Wilson gauge action~\cite{Kitazawa:2019otp}. Various $L_\tau=N_\tau a$ and $L_x=N_x a$ are simulated on the lattices of size 
$N_x\times N_z^2\times N_\tau$ with the PBC for all directions and $a$ being the lattice spacing. We performed simulations for $N_\tau=12,16$ to study discretization effects. The lengths along the other two directions, $N_z$, are taken as $N_z/N_\tau\ge6$ to suppress the effects of BCs along these directions.
To measure thermodynamic quantities, we use the expectation values of the energy-momentum tensor operator introduced by the small flow-time expansion method~\cite{Suzuki:2013gza,Kitazawa:2016dsl,Iritani:2018idk}. 

\begin{figure}
    \centering
\includegraphics[width=0.6\linewidth]{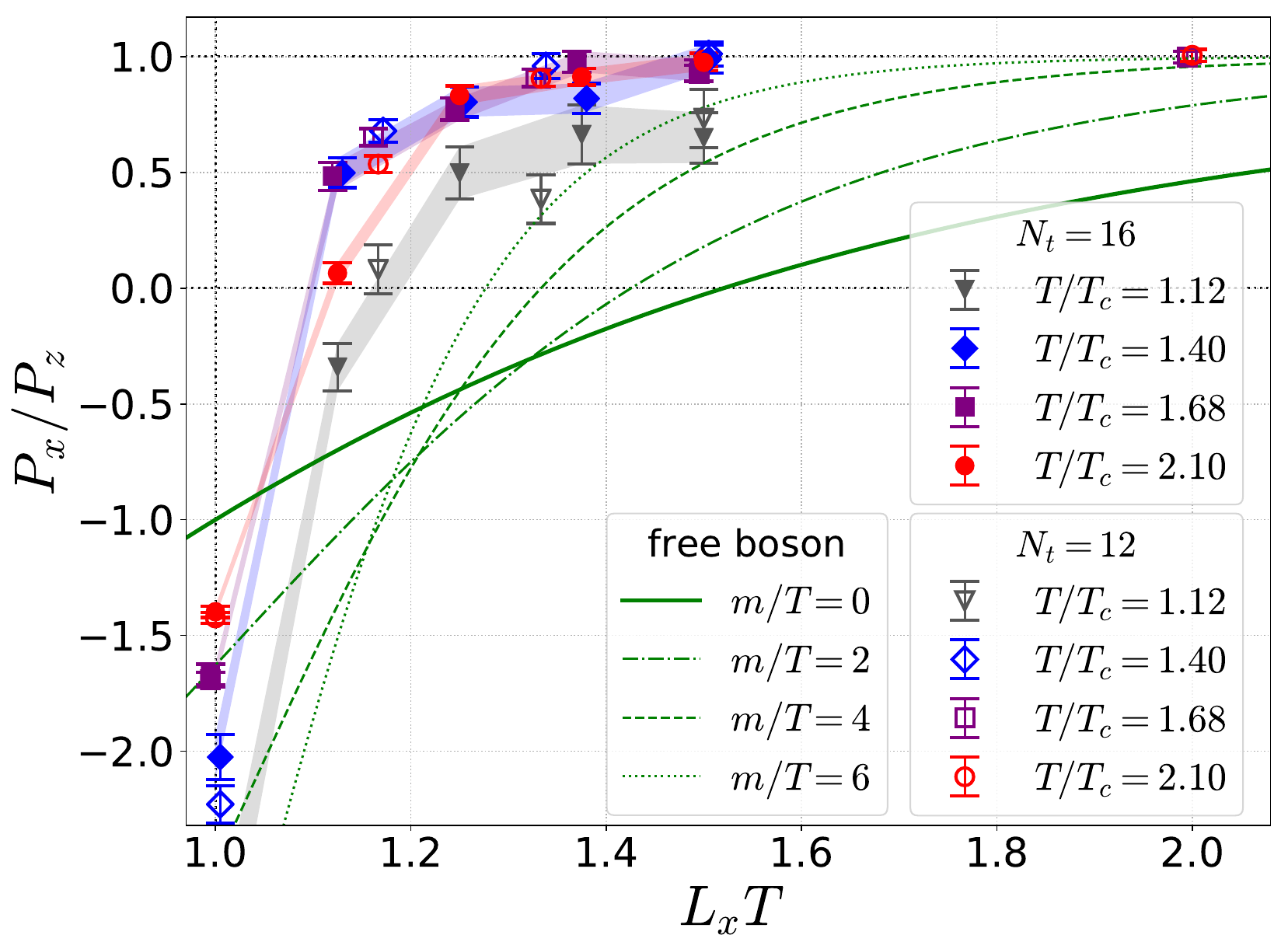}
    \caption{Pressure ratio $P_x/P_z$ as a function of $L_xT$ for various values of $T/T_c$ and $N_t=16,12$~\cite{Kitazawa:2019otp}. The behavior of $P_x/P_z$ in the free scalar theory is also shown by the lines for several values of mass-temperature ratio $m/T$.}
    \label{fig:lat_ratio}
\end{figure}

In Fig.~\ref{fig:lat_ratio}, we plot the numerical results on the ratio of the pressure along $x$ and $z$ directions, $P_x=\langle T_{11}\rangle$ and $P_z=\langle T_{22}\rangle=\langle T_{33}\rangle$ as a function of $L_xT$ for four temperatures above $T_c$ by the points with errors~\cite{Kitazawa:2019otp}. We also show the behavior of $P_x/P_z$ in the free-boson system of mass $m$ by the green lines. The ratio $P_x/P_z$ should become unity for $L_xT\to\infty$ because the full rotational symmetry is restored in this limit. In the free-boson result, a clear deviation from this limit is observed already at $L_xT=2$. However, the lattice results show $P_x/P_z = 1$ within statistics even at $L_xT = 1.5$ for $1.4\le T/T_c\le2.1$. Even at $L_xT = 11/8=1.375$ and $4/3=1.333$, deviation from $P_x/P_z = 1$ is comparable with statistical errors. As $L_xT$ is decreased further, the ratio suddenly becomes small and arrives at negative values $P_x/P_z < -1$ at $L_xT = 1$. It is notable that this behavior is observed in a wide range of temperature for $1.4\le T/T_c\le2.1$, while the result at $T /T_c = 1.12$ has a deviation from this trend. From these results, we conclude that the SU(3) YM theory at $1.4 \le T /T_c \le 2.1$ is remarkably insensitive to the PBC along a spatial direction compared with the massless free theory. 

At asymptotically high temperatures, YM theories approach a free gas composed of massless gluons. The $L_xT$ dependence of $P_x/P_z$ in this limit thus should approach that of the massless free scalar theory. It is an interesting question how the thermodynamics on $\TTRR$ approach this
asymptotic limit. However, an extension of the lattice study to high $T$ has two difficulties. First,
for finer lattice spacing $a=(N_\tau T)^{-1}$, the simulations on larger lattices are required for the vacuum subtraction. Second, the relation between the gauge coupling and $a$ is unknown for fine lattices. To avoid these difficulties, in the following we focus on the ratio
\begin{align}
    R = \frac{P_x+\delta}{P_z+\delta} ,
    \label{eq:delta}
\end{align}
with $\delta= (\epsilon-P_x-2P_z)/4$ with the energy density $\epsilon$. Equation~\eqref{eq:delta} can be analyzed straightforwardly on the lattice even for small $a$~\cite{Kitazawa:2019otp}. 

\begin{figure}
    \centering
\includegraphics[width=0.48\linewidth]{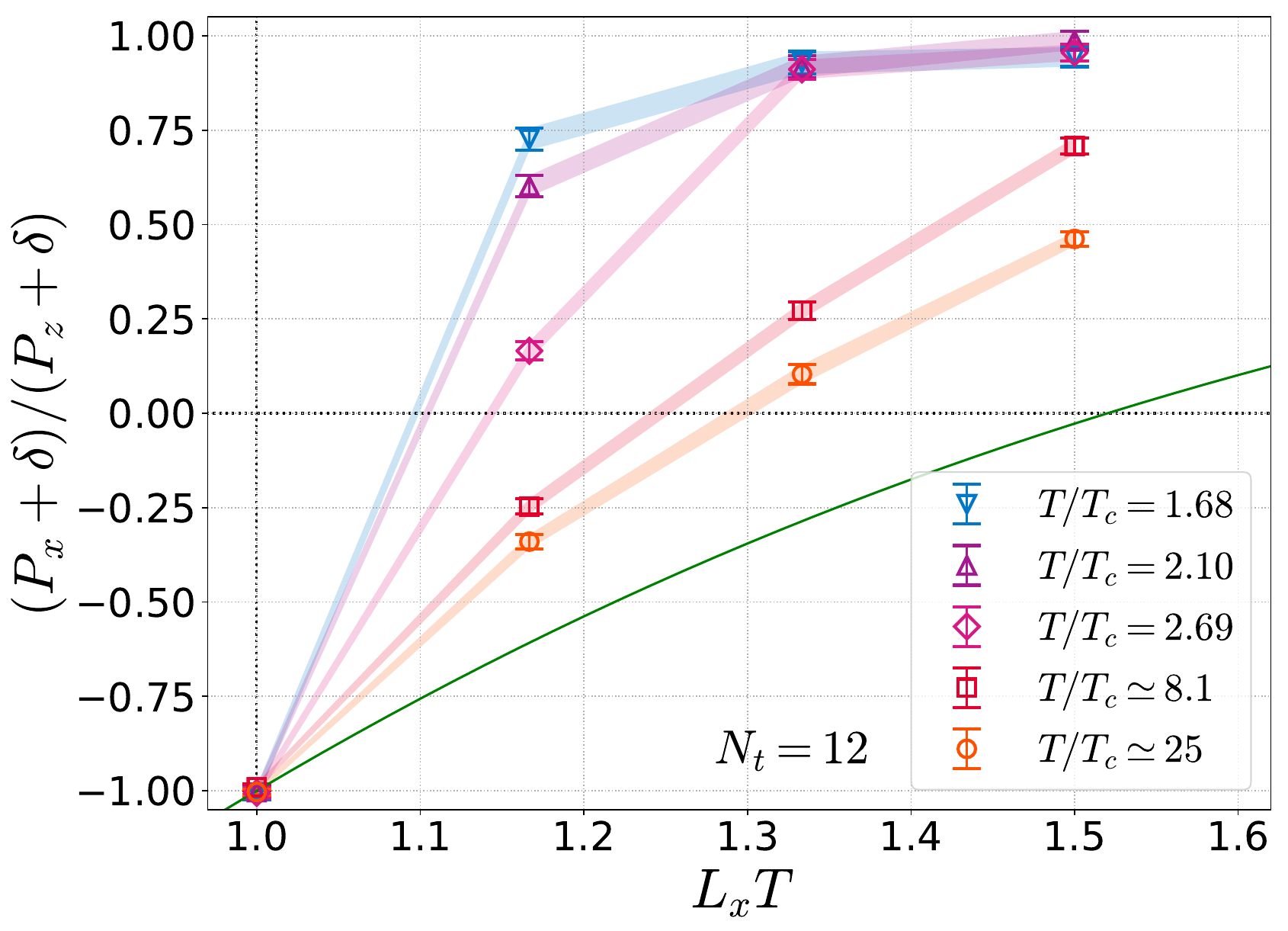}
\includegraphics[width=0.48\linewidth]{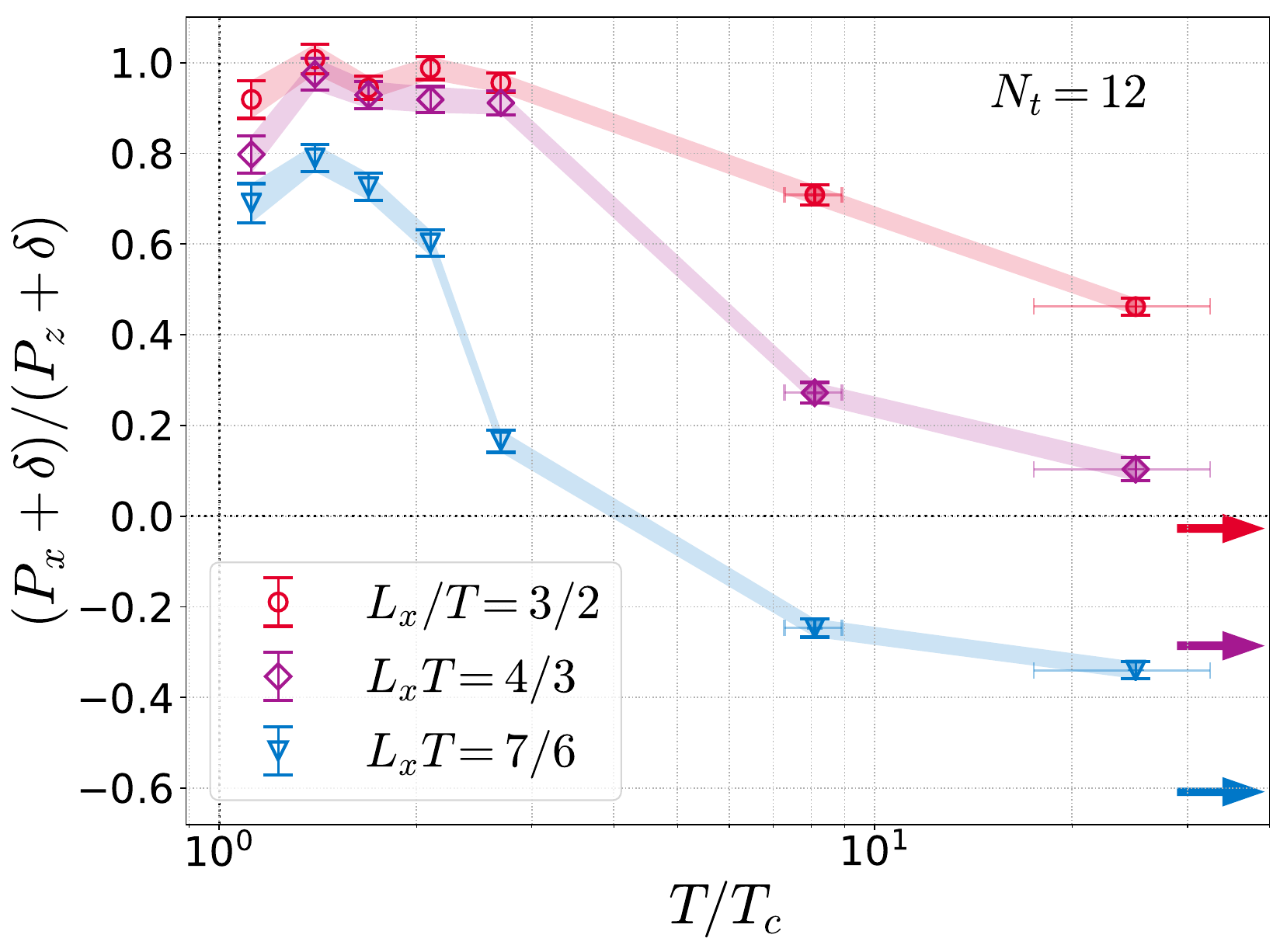}
    \caption{Ratio $(P_x + \delta)/(P_z + \delta)$ for various values of $T$ and $L_xT$. The left (right) panel shows the ratio as a function of $L_xT$ ($T /T_c$). The solid line in the left panel shows the ratio in the massless free scalar theory. The arrows in the right panel show the ratio in the massless free scalar theory for each $L_xT$.}
    \label{fig:lat_highT}
\end{figure}

In Fig.~\ref{fig:lat_highT}, we show the behavior of Eq.~\eqref{eq:delta} as functions of $L_xT$ and $T /T_c$ in the left and right panels, respectively. In the left panel, the ratio~\eqref{eq:delta} in the massless free scalar theory is depicted by the solid-green line, while
in the right panel the ratio for each $L_xT$ is shown by the arrows. The comparison of the lattice data with these results suggests that the former approaches the asymptotic value as $T$ is increased, but the difference is still large even at $T/T_c=25$.
In this way, the lattice results show a peculiar behavior of thermodynamics on $\TTRR$. 

\section{Model Study}
\label{sec:model}

To explore the physical origin behind the lattice results discussed above, now we study them in an effective model~\cite{Suenaga:2022rjk,Fujii:2024llh}. 

\subsection{Model construction}

To introduce an effective model that describes YM theory on $\TTRR$, we focus on the fact that there are two $Z_N$ symmetries associated with two PBCs in $\tau$ and $x$ directions that can be spontaneously broken with the variation of $L_\tau$ and $L_x$. This property indicates that the effective model should be able to describe these spontaneous symmetry breakings simultaneously.

We construct such a model based on an effective model proposed in Ref.~\cite{Meisinger:2001cq}, which is the model describing the thermodynamics of $SU(N)$ YM theory on $\SRRR$. The key ingredient of this model is the Polyakov loop $\Omega_\tau$, which is an order parameter of $Z_N$ symmetry associated with the PBC in the temporal direction. The expectation value of $\Omega_\tau$ is determined to minimize the free-energy density, which consists of the perturbative and potential terms as 
\begin{align}
    f^{\SRRR}(\vec\theta_\tau;L_\tau)
    =f_{\rm pert}^{\SRRR}(\vec\theta_\tau;L_\tau)+f_{\rm pot}^{\SRRR}(\vec\theta_\tau;L_\tau),
    \label{ftot}
\end{align}
where $\vec\theta_\tau$ is the three-vector representing the phase of three eigenvalues of the Polyakov-loop matrix. Here, the perturbative term $f_{\rm pert}^{\SRRR}$ represents the 
free energy of non-interacting gluons with the uniform background field corresponding to the nonzero expectation value of $\Omega_\tau$. The potential term $f_{\rm pot}^{\SRRR}$ is the phenomenological term introduced to reproduce confinement phase transition at $T=T_c$ and lattice data on thermodynamics. In Ref.~\cite{Meisinger:2001cq}, it was found that this simple model is capable of reproducing the qualitative behavior of the lattice data. This idea has been later elaborated in Ref.~\cite{Dumitru:2012fw} for better reproduction of the lattice data. 

In the present study, we extend the model in Ref.~\cite{Dumitru:2012fw} to $\TTRR$ by introducing two Polyakov loops as dynamical degrees of freedom; 
the Polyakov loop along $x$ direction, $\Omega_x$, is introduced in addition to the conventional temporal loop $\Omega_\tau$. The free-energy density read
\begin{align}
    f(\vec\theta_\tau,\vec\theta_x;L_\tau,L_x)
    = f_{\rm pert}(\vec\theta_\tau,\vec\theta_x;L_\tau,L_x)
    + f_{\rm pot}(\vec\theta_\tau,\vec\theta_x;L_\tau,L_x),
    \label{eq:f}
\end{align}
where $\vec\theta_x$ represents three eigenvalues in $\Omega_x$. For $f_{\rm pot}$, we introduce an ansatz
\begin{align}
    f_{\rm pot}(\vec{\theta}_\tau,\vec{\theta}_x;L_\tau,L_x)
    =& \, 
    f_{\rm sep}(\vec{\theta}_\tau,\vec{\theta}_x;L_\tau,L_x)
    +f_{\rm cross}(\vec{\theta}_\tau,\vec{\theta}_x;L_\tau,L_x), 
    \label{FPotSeparate}
    \\
    f_{\rm sep}(\vec{\theta}_\tau,\vec{\theta}_x;L_\tau,L_x)
    =& \,
    f^{\SRRR}_{\rm pot}(\vec{\theta}_\tau,L_\tau)
    +f^{\SRRR}_{\rm pot}(\vec{\theta}_x,L_x),
    \label{eq:fsep}
\end{align}
where we use $f^{\SRRR}_{\rm pot}$ determined in Ref.~\cite{Dumitru:2012fw}. The cross term $f_{\rm cross}$ consists of the terms containing both $\Omega_\tau$ and $\Omega_x$, and hence represents their interplay. In Ref.~\cite{Suenaga:2022rjk}, it was found that the introduction of $f_{\rm cross}$ is essential in reproducing the lattice data.

In the present study, the form of the cross term $f_{\rm cross}$ is determined in a phenomenological manner within the constraints from the symmetries of the system and limiting behaviors for $L_{\tau,x}\to0$ and $\infty$; see Ref.~\cite{Fujii:2024llh} for detailed arguments and its explicit form. The form of $f_{\rm cross}$ employed here contains four free parameters, which are determined so as to reproduce the lattice results in Ref.~\cite{Kitazawa:2019otp}. 

In the following, we denote the critical temperature of the deconfinement phase transition on $\SRRR$ as $T_{\rm d}$.

\subsection{Numerical results}

\begin{figure}
    \centering
\includegraphics[width=0.6\linewidth]{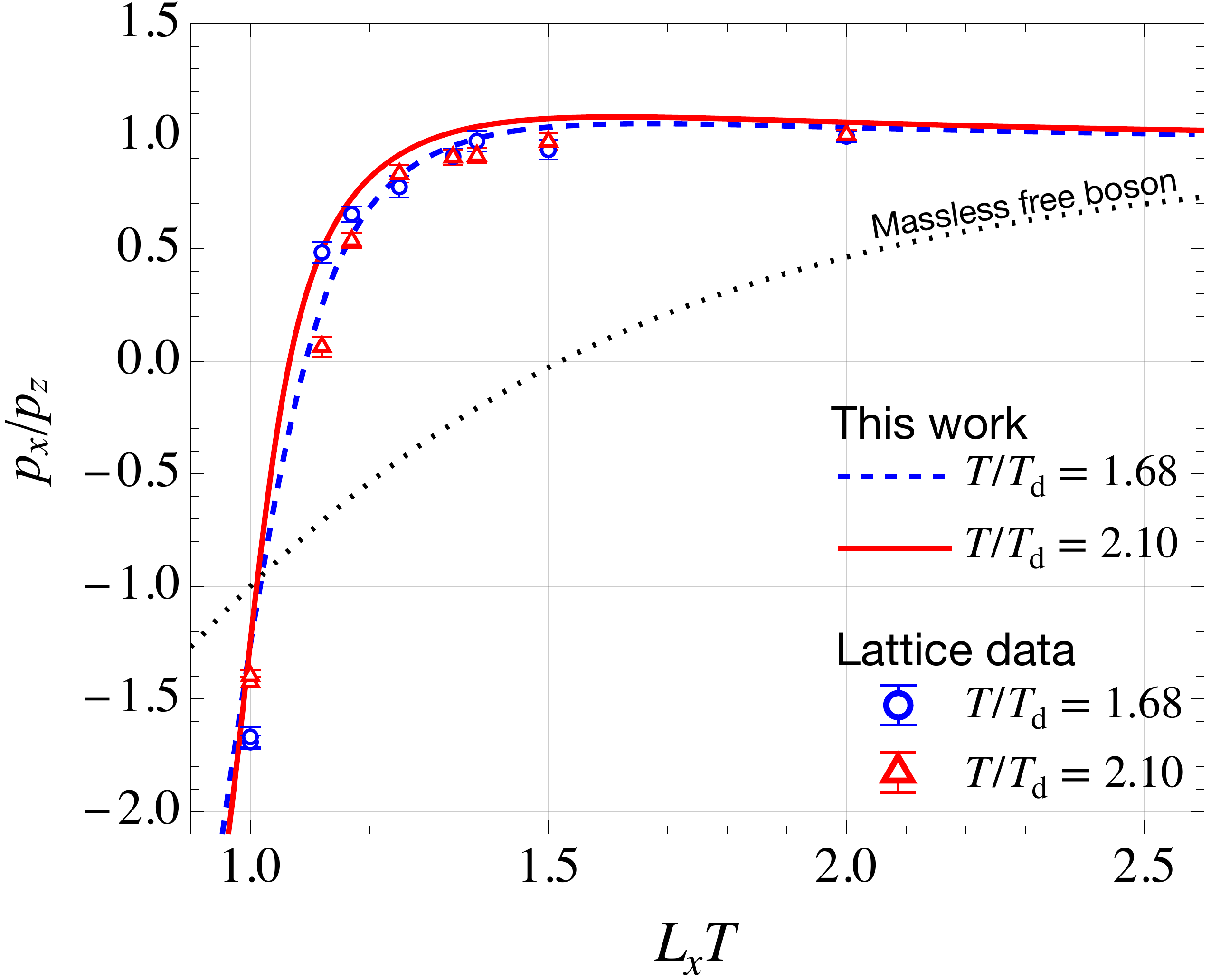}
    \caption{$L_xT$ dependence of the pressure ratio $p_x/p_z$ at $T /T_{\rm d} = 1.68$ and $2.10$, together with the lattice data in Sec.~\ref{sec:lattice}~\cite{Fujii:2024llh}.}
    \label{fig:model_ratio}
\end{figure}

In Fig.~\ref{fig:model_ratio}, we first show the $L_x T$ dependence of the pressure ratio $p_x/p_z$ at $T/T_{\rm d}=1.68$ and $2.10$ by the blue-dashed and red-solid lines. The lattice data in Ref.~\cite{Kitazawa:2019otp} are also indicated by discrete points with errorbars~\cite{Fujii:2024llh}. The $L_xT$ dependence of $p_x/p_z$ in the massless-free system is also shown by the dotted line. 
From the figure one finds that the lattice results are qualitatively reproduced by the model analysis. Our analyses also show that our model well reproduces individual thermodynamic quantities besides the ratio $p_x/p_z$ for these temperatures~\cite{Fujii:2024llh}.

\begin{figure}
    \centering
\includegraphics[width=0.9\linewidth]{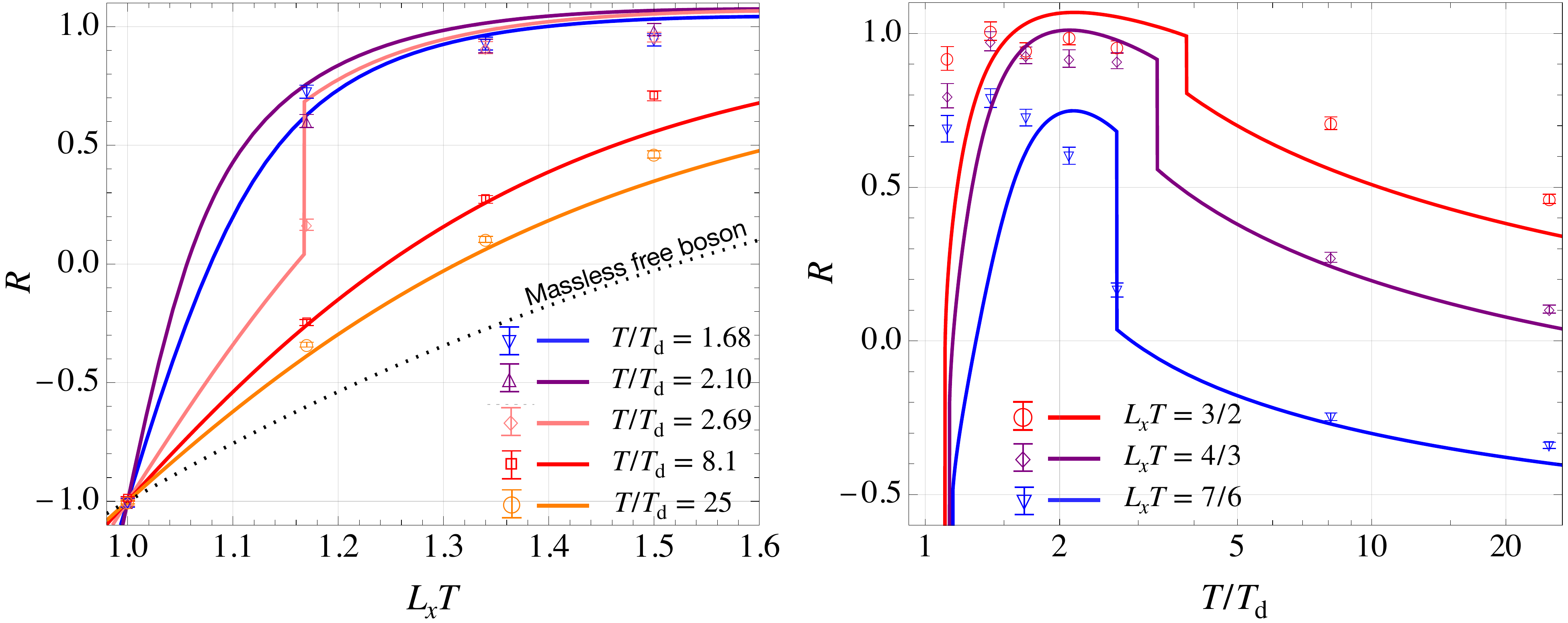}
    \caption{Left: $L_xT$ dependence of the ratio $R=(p_x+\delta/4)/(p_z+\delta/4)$ in Eq.~\eqref{eq:delta} for several values of $T/T_{\rm d}$. Right: $T/T_{\rm d}$ dependence of $R$ for various $L_xT$.}
    \label{fig:model_highT}
\end{figure}

Next, in Fig.~\ref{fig:model_highT} we compare the behavior of Eq.~\eqref{eq:delta} in the model analysis with the lattice data. The figure shows that the lattice results are well reproduced by the model, while the agreement becomes worth for lower $T$. Another interesting feature of the model analysis is that it has a discontinuous jump, i.e. a first-order phase transition, at $T/T_{\rm d}=2.69$ and $L_xT\simeq 1.17$.
This first-order phase transition is more clearly seen in the right panel. Unfortunately, the current lattice data in Ref.~\cite{Kitazawa:2019otp} is not sufficiently fine to verify the existence of the discontinuity. However, the lattice data at $L_xT=4/3,~7/6$ change rapidly around $T/T_{\rm d}\simeq3$, which may support the existence of the first-order transition.

\begin{figure}
    \centering
\includegraphics[width=0.43\linewidth]{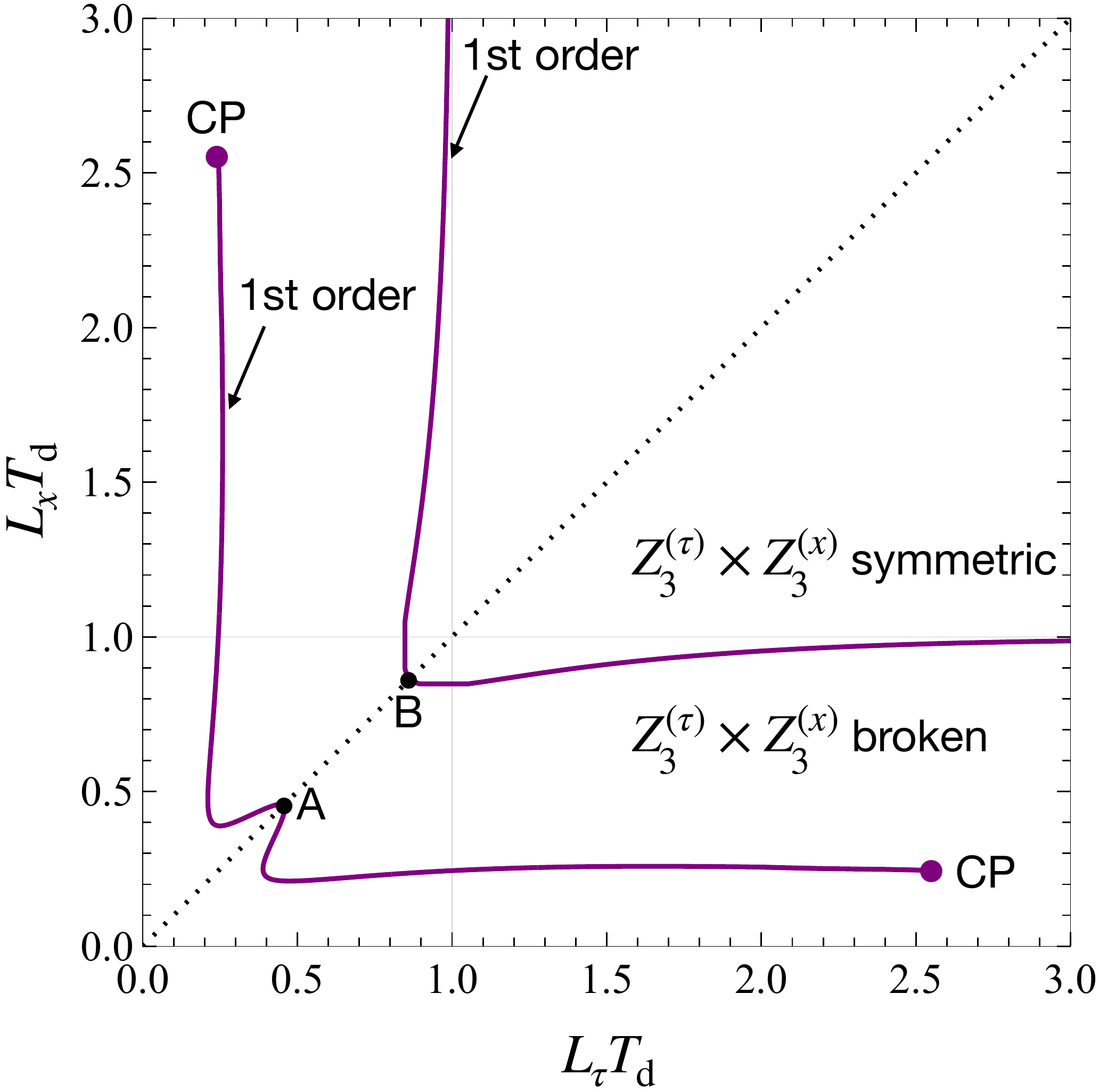}
\hspace{3mm}
\includegraphics[width=0.47\linewidth]{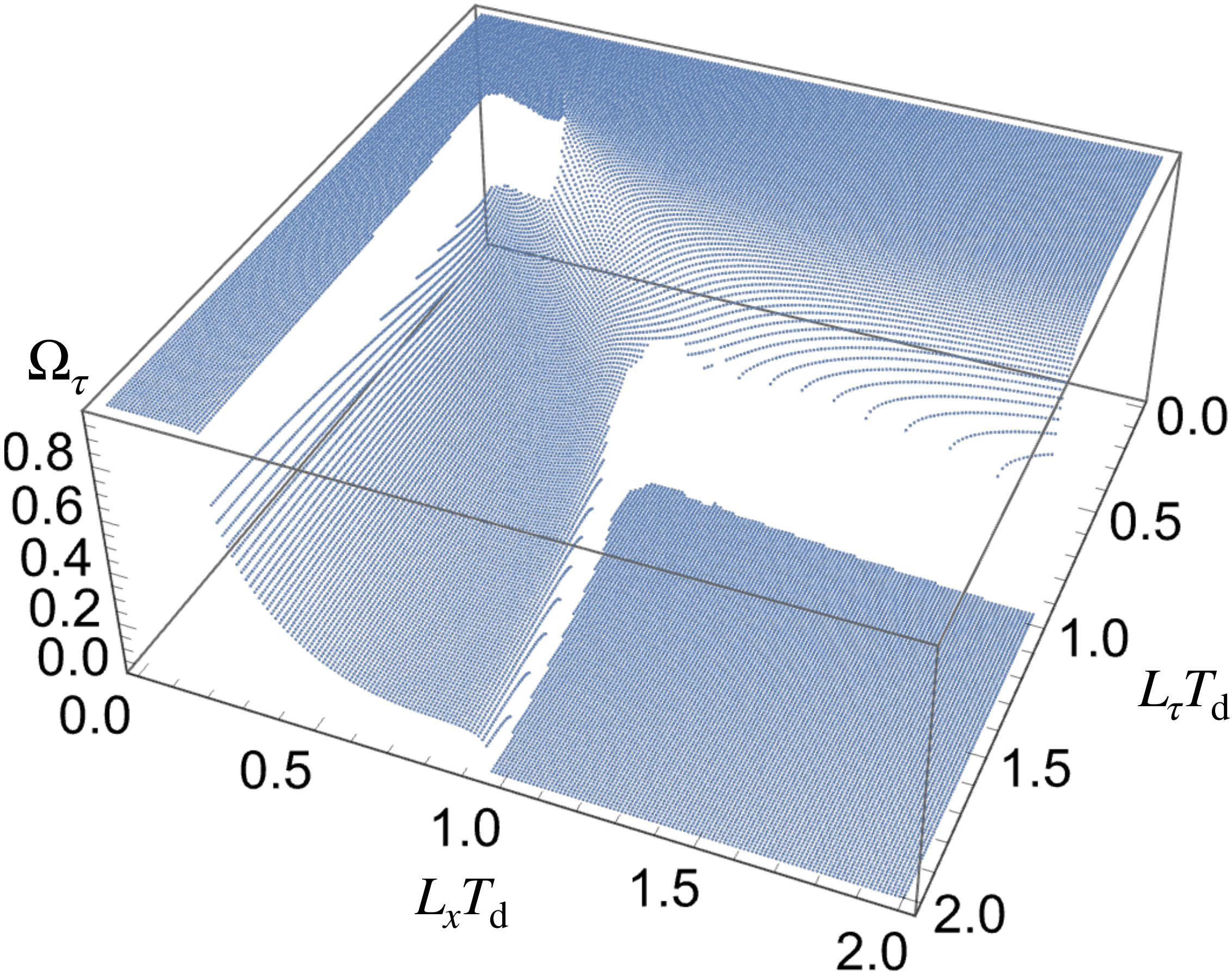}
    \caption{Left: Phase diagram on the $L_\tau$--$L_x$ plane, where the first-order phase transitions are shown by the solid lines. Right: Behavior of $\Omega_\tau$ as a function of $L_\tau$ and $L_x$.}
    \label{fig:phased}
\end{figure}

In the left panel of Fig.~\ref{fig:phased}, we plot the phase diagram on the $L_\tau$--$L_x$ plane obtained in our model. The solid lines show the first-order phase transitions at which the thermodynamic quantities change discontinuously. The phase diagram is symmetric with the exchange of $L_\tau$ and $L_x$  because the system is invariant under the exchange of $\tau$ and $x$ axes. In the right panel of Fig.~\ref{fig:phased}, we also show the behavior of $\Omega_\tau$ on the $L_\tau$--$L_x$ plane. 

Figure~\ref{fig:phased} shows that there are two first-order transition lines on the $L_\tau$--$L_x$ plane, and two Polyakov loops, $\Omega_\tau$ and $\Omega_x$, change discontinuously there. Among them, the line including point~B in the left panel is connected to the confinement phase transition on $\SRRR$ in the large $L_x$ limit at $L_\tau=1/T_{\rm d}$. From the right panel, one finds that the upper-right region of this line is the confined phase, where both $Z_3$ symmetries are restored with $\Omega_\tau=\Omega_x=0$. On the other hand, both $Z_3$ symmetries are spontaneously broken in the lower-left region of the line with $\Omega_\tau\ne0$ and $\Omega_x\ne0$. The result of the model analysis does not have the phases where only one of the $Z_3$ symmetries is spontaneously broken. 

The other first-order transition line on the phase diagram including point~A corresponds to the one found in Fig.~\ref{fig:model_highT}.
As seen from the left panel, this transition line lies entirely in the $Z_3$ broken phase. Furthermore, this line terminates at finite $L_\tau$ and $L_x$ at 
$(L_\tau T_{\rm d},L_x T_{\rm d})\simeq(0.25,2.54)$ 
and $(2.54,0.25)$, and hence is not connected to phase transitions on $\SRRR$.

An end point of a first-order transition line is the critical point at which the phase transition is of second order. The universality class of these critical points should be the same as that in the two-dimensional Ising model for the following reasons. First, on the first-order phase transition including point~A, two phases characterized by different values of order parameters coexist. Second, the correlation length grows near the critical points and eventually exceeds $L_\tau,\ L_x$. Then, the system should be regarded as two-dimensional. 

\section{Summary}

In this proceeding, 
we studied the thermodynamics and phase structure of $SU(3)$ YM theory on $\TTRR$ in Euclidean spacetime, which corresponds to a thermal system with a PBC along one spatial direction, in lattice Monte Carlo simulations and an effective model. In lattice simulations, we found that a clear pressure anisotropy is observed only at a significantly shorter $L_x$ compared with the free scalar theory. We then investigated the thermodynamics obtained on the lattice in an effective model that incorporates two Polyakov loops along two compactified directions. The free parameters in the model have been determined to reproduce thermodynamics measured on the lattice. We found that the interplay between two Polyakov loops plays an essential role in reproducing the lattice data. The model analysis indicates the existence of a novel first-order phase transition and critical points. It is an interesting future study to compare these results with the analysis of $\TTRR$ in various systems~\cite{Hanada:2010kt,Unsal:2010qh,Mandal:2011hb,Ishikawa:2019dcn,Chernodub:2022izt,Hayashi:2024qkm,Fujii:2024fzy,Fujii:2024ixq,Fujii:2024woy}.

\subsection*{Acknowledgements}

This work was supported in part by JSPS KAKENHI (Nos. JP22K03619, JP23K03377, JP23H04507, JP23H05439, JP24K07049, JP24K17054),
the Center for Gravitational Physics and Quantum Information (CGPQI) at Yukawa Institute for
Theoretical Physics.

\bibliographystyle{JHEP}
\bibliography{reference}

\end{document}